\documentclass[useAMS,usenatbib]{mn2e}
\usepackage{graphicx}
\newcommand{\etal}{{\it et al.} }

\newcommand{\fuse}{{\it FUSE} }
\newcommand{\xmm}{{\it XMM-Newton} }
\newcommand{\chandra}{{\it Chandra} }

\newcommand{\hetg}{{\it HETGS} }

\newcommand{\nseven}{N~{\sc vii} }
\newcommand{\neten}{Ne~{\sc x} }
\newcommand{\mgtwelve}{Mg~{\sc xii} }
\newcommand{\sifourteen}{Si~{\sc xiv} }

\newcommand{\silya}{Si~{\sc xiv}~Ly$\alpha$ }
\newcommand{\nelya}{Ne~{\sc x}~Ly$\alpha$ }

\newcommand{\mglya}{Mg~{\sc xii}~Ly$\alpha$ }

\newcommand{\oxysix}{O~{\sc vi} } 
\newcommand{\oxyseven}{O~{\sc vii} }
\newcommand{\oxyeight}{O~{\sc viii} }
\newcommand{\oxynine}{O~{\sc ix} }

\newcommand{\nenine}{Ne~{\sc ix} }

\newcommand{\neniner}{Ne~{\sc ix} (r) }

\newcommand{\mgeleven}{Mg~{\sc xi} }

\newcommand{\sithirteen}{Si~{\sc xiii} }

\newcommand{\thr}{3C~120 }

\newcommand{\mcg}{MCG~$-$6$-$30$-$15 }

\title[Constraints on hot metals in the Vicinity of the Galaxy]{CONSTRAINTS ON HOT METALS IN THE VICINITY OF THE GALAXY}

\author[B. McKernan, T. Yaqoob \& C. S. Reynolds]{B. McKernan$^{1}$\thanks{E-mail:mckernan@astro.umd.edu (BMcK)}, T. Yaqoob$^{2,3}$ and C. S. Reynolds$^{1}$\\
$^{1}$Department of Astronomy, University of Maryland, 
College Park, MD 20742\\
$^{2}$Department of Physics and Astronomy,
                        Johns Hopkins University, Baltimore, MD 21218\\
$^{3}$Laboratory for High Energy Astrophysics,
                NASA/Goddard Space Flight Center, Greenbelt, MD 20771}
\begin{document}

\date{Accepted. Received; in original form}

\pagerange{\pageref{firstpage}--\pageref{lastpage}} \pubyear{2004}

\maketitle

\label{firstpage}

\begin{abstract}
We have searched for evidence of soft X-ray absorption by hot metals in the 
vicinity of the Galaxy in the spectra of a small sample of fifteen type~I AGN 
observed with the high resolution X-ray gratings on board 
\emph{Chandra}. This is an extension of our previous survey of hot 
\oxyseven and \oxyeight absorbing gas in the vicinity of the Galaxy. The 
strongest absorption signatures within a few hundred km $\rm{s}^{-1}$ of 
their rest-frame energies are most likely due to warm absorbing outflows 
from the nearest AGN, which are back-lighting the local hot gas. We 
emphasize that absorption signatures in the spectra of some distant AGN 
that are kinematically consistent with the recessional velocity of the 
AGN are most likely to be due to hot local gas. Along the sightline 
towards PG~1211$+$143, PDS~456 and MCG-6-30-15 there is a very large 
absorbing Fe column density which is kinematically consistent with 
absorption by hot, local Fe. The sightlines to these three AGN pass through 
the limb of the Northern Polar Spur (NPS), a local bubble formed from several 
supernovae which, if rich in Fe, may account for a large local Fe column.

We obtain limits on the column density of local, highly ionized N, Ne, Mg, Si 
along all of the sightlines in our sample. We correlate the column density 
limits with those of highly ionized O along the same sightlines. Assuming 
the hot local gas is in collisionally ionized equilibrium, we obtain 
limits on the temperature and relative abundances of the metals in the 
hot local gas. Our limits on the ionic column densities in the local 
hot gas seem to be consistent with those observed in the hot halo gas 
of edge-on normal spiral galaxies. 

\end{abstract}


\section{Introduction}
\label{sec:intro}
Present-day structures such as galaxies and clusters of 
galaxies are believed to have condensed from within partially collapsing 
filaments of primordial matter. The legacy of the structure formation 
epoch includes shock-heated filaments as well as structures and these 
filaments should thread the modern universe. Simulations of structure 
formation indicate that around half of the baryonic 
matter at low-redshift lives in such filaments, shock-heated to 
$\sim 10^{5-7}$K (see e.g. \citet{b4}; \citet{b6} and references 
therein). Although more recent work by \citet{b41} suggests that a 
a lower temperature component (T$<10^{5}$ K) of the shock-heated 
filaments may be more significant than previously thought. The `warm' 
($\sim 10^{5-6}$K) component of this warm/hot intergalactic medium 
(WHIGM) has been observed in absorption in the UV band (see e.g. 
\citet{b22} and references therein). However, much of the WHIGM is 
expected to be hotter than this, so the high spectral resolution 
X-ray detectors such as those aboard \chandra and \xmm are best 
placed for investigating the `hot' component of the missing baryons. 

The clearest X-ray spectral signature of hot gas in the vicinity of our 
Galaxy consists of absorption features imprinted in spectra of 
X-ray bright active galactic nuclei (AGN) at $z=0$ in the observed 
frame \citep{b10,b18}.  One of the most important new results from the 
\chandra and {\it XMM-Newton} X-ray telescopes, has been the 
discovery of hot, low density, highly ionized gas in the vicinity of 
our Galaxy (at $z=0$) and possibly beyond (at $z>0$). X-ray absorption 
due to local ($z=0$) hot gas was discovered recently 
\citep{b17,b8,b2,b19,b9,b12,b13,b14}. 
Of course, such absorption need not be due to local WHIGM. 
Absorption by hot gas at $z=0$ could be due to local 
Galactic gaseous structures or infalling High Velocity Clouds 
(see e.g. \citet{b20}; \citet{b34} and references therein). Even if 
there are no known local structures along the sightline 
to a particular AGN, the location of the absorbing gas is still ambiguous. 
Around half of all type~I AGN exhibit strong 
absorption in the soft X-ray and UV bands due to partially ionized, 
optically thin, outflowing, circum-nuclear material known as the 
`warm absorber' (see e.g. \citet{b15} and references therein). 
Since the warm absorbers are outflowing, absorption 
which is actually local to our Galaxy at $z \sim 0$ could be 
misinterpreted as a warm absorber outflow coinciding with the cosmological 
recession velocity (cz) of the AGN (see e.g. Fig.3 of \citet{b14}). Confusion
 with AGN outflow is a worse problem for WHIGM at intermediate redshifts 
($z>0$). The few apparently robust detections of absorbers at redshifts 
intermediate between our Galaxy and the AGN rest-frame, e.g. \citet{b23} 
could be variable (\citet{b30}) suggestive of an origin in an ionized 
outflow from the host AGN, rather than WHIGM.

In this paper, we investigate a sample of AGN observed with the high energy
 transmission gratings \citep{b11} on board \chandra. The 
uniform analysis of the data from these AGN and the results of the analysis,
 in particular the characterization of the AGN continua and the warm 
absorption in AGN, have been discussed in detail by \citep{b15}. 
In \citet{b14} we investigated this sample of AGN spectra for
 absorption due to local, highly ionized Oxygen. Here we extend that study 
by searching for absorption in the vicinity of the Galaxy by metals 
other than Oxygen. Our aim is to further the systematic study of the hot 
local gas and to begin constraining the temperatures and the 
relative metal abundances in the hot, local gas.

\section{The Sample and Data Analysis}
\label{sec:obs}
 
Table~\ref{tab:1} lists the AGN sample assembled by \citet{b15}. 
Also listed in Table~\ref{tab:1} are the AGN redshifts 
(from NED\footnote{http://nedwww.ipac.caltech.edu/forms/byname.html} using 
21cm H{\sc~i} radiation measurements where possible), the AGN Galactic 
latitude and longitude (also from NED), the Galactic column density and 
the total exposure times of the spectra. The sample, including selection 
criteria are discussed in detail in \citet{b15}. The \chandra 
data were reprocessed and analyzed according to the methods outlined in 
\citep{b15}.

\begin{table}
\begin{minipage}{85mm}
\caption{The \chandra \hetg sample of type~I AGN \label{tab:1}.
 Columns 2,3 and 4 give the Galactic co-ordinates and the 
redshift of the source (from  NED). Redshift
       was deduced from observations of the 21cm H~{\sc{i}} line where
      possible, since  optical estimates of $z$ may be confused by AGN
 outflow. $^{a}$ Galactic column density from \citet{b7}, except for Mkn
     509 \citep{b16}.  The Galactic column density towards F9,
           NGC~3227, NGC~3516, NGC~3783, NGC~5548, Mkn 766, NGC~7314 
and  Akn 564 was estimated from
   interpolations from the measurements of  Stark \etal (1992). $^{b}$ 
\thr is also classified as a broad-line radio galaxy (NED).}
\begin{tabular}{@{}lrrrc@{}}
\hline
Source & Gal. long. & Gal. lat. 
&Redshift & Gal. $N_{H}$ \\
 & ($^{\circ}$) & ($^{\circ}$) & (z) & 
($10^{20} \rm{cm}^{-2}$) $^{a}$ \\
\hline
  Fairall 9 & 295.07 & -57.83 & 0.04600 & 3.0\\  
3C 120$^{b}$ & 190.37 & -27.40 & 0.03301 & 12.30 \\ 
NGC~3227 &216.99 &55.45 &0.00386 &2.15 \\
NGC~3516 &133.24 &44.40 &0.00884 & 3.05 \\
NGC 3783 & 287.46 & 22.95 & 0.00973 &8.50 \\ 
NGC 4051 & 148.88 & 70.09 & 0.00242 & 1.31\\ 
Mkn 766 & 190.68 & 82.27 & 0.01293 & 1.80\\ 
NGC 4593 &297.48 & 57.40 & 0.00831 & 1.97\\ 
MCG-6-30-15 & 313.29& 27.68 & 0.00775&4.06\\ 
IC 4329a & 317.50& 30.92 &0.01605 & 4.55 \\
Mkn~279 &115.04 &46.86 &0.03045 & 1.64 \\ 
NGC 5548 & 31.96 & 70.50 & 0.01717 & 1.70 \\ 
Mkn 509 & 35.97 & -29.86 & 0.03440 & 4.44 \\   
NGC~7314 &27.14 &-59.74 &0.00474 & 1.46 \\
Akn 564 &92.14 & -25.34 & 0.02467 & 6.40 \\
\hline
\end{tabular}
\end{minipage}
\end{table}

Here we extend the study of \citep{b14} by searching for evidence of 
absorption by metals less abundant than Oxygen. Table~\ref{tab:2} 
lists the relative solar abundances of the metals relevant for this study. 
Iron is the next 
most abundant metal after those listed in Table~\ref{tab:2}, but 
there is a forest of Fe L- and M-shell transitions in the 
soft X-ray band (see e.g. \citet{b40}). There is therefore considerable 
ambiguity in Fe absorption line identification due to blending. Furthermore, 
higher order transitions of more abundant elements can also be 
mis-identified as 
Fe transitions. Therefore we limited our search to the strongest absorption
 transitions in the soft X-ray band in the most abundant highly stripped 
ions (not \oxyseven and \oxyeight since we have studied these elsewhere) 
namely: \nseven, \nenine, \neten, \mgeleven, \mgtwelve, 
\sithirteen and \sifourteen respectively. Table~\ref{tab:3} lists the 
transitions that we investigated in this study, 
including their rest-frame wavelengths and oscillator strengths. 

\begin{table}
\begin{center}
\begin{minipage}{65mm}
\caption{Relative solar element abundances relevant 
for this study. Values are taken from solar photospheric 
values of \citet{b1}. \citet{b31} suggest that the correct 
solar abundance of Ne is [Ne/H]$=8.29\pm 0.05$, values 
in (brackets) use this value of the solar Ne abundance. 
\label{tab:2}}
\begin{tabular}{@{}lrrr@{}}
\hline
X & $\log(X/O)_{\odot}$&$\log(X/Ne)_{\odot}$ &$\log(X/Si)_{\odot}$\\
\hline
O & 0.00 & 0.84(0.64) & 1.38\\  
Ne &-0.84(-0.64) &0.00 & 0.54(0.74)\\  
N &-0.88 &-0.04(-0.24) & 0.50\\  
Mg &-1.34 &-0.50(-0.70) & 0.04\\  
Si & -1.38& -0.54(-0.74)& 0.00\\  
\hline
\end{tabular}
\end{minipage}
\end{center}
\end{table}

\begin{table}
\begin{center}
\begin{minipage}{55mm}
\caption{Details of the discrete absorption transitions discussed in 
this study. \label{tab:3} f denotes the oscillator strength and 
$\lambda$ the wavelength (in \AA\ ) of the 
respective transitions (values from the Atomic Line List at 
http://www.pa.uky.edu/\~\ peter/atomic).}
\begin{tabular}{@{}lcrr@{}}
\hline
Ion & Transition & f & $\lambda$(\AA\ )\\
\hline
\nseven &Ly$\alpha$ &0.416 & 24.7810\\  
\nenine &(r)${\it 1s-2p}$ &0.724 & 13.4473 \\ 
\neten &Ly$\alpha$&0.415 & 12.1339 \\
\mgeleven& (r)${\it  1s-2p}$&0.742 &9.1688 \\
\mgtwelve&Ly$\alpha$ & 0.415 & 8.4210 \\ 
\sithirteen& (r)${\it  1s-2p}$& 0.757 & 6.6480\\ 
\sifourteen&Ly$\alpha$ & 0.414 & 6.1822\\ 
\hline
\end{tabular}
\end{minipage}
\end{center}
\end{table}

Once we measured the discrete absorption profiles, we used the 
extrapolated linear approximation to the curves-of-growth\footnote{
The linear part of the curves-of-growth implies that 
$N_{ion}=1.13 \times 10^{17} EW/ f \lambda^{2}$ 
where $N_{ion}$ is the ionic column density ($\rm{cm}^{-2}$), EW is the 
equivalent width of the absorption feature (in m\AA\ ), 
f is the oscillator strength of the transition and $\lambda$ is in \AA .}  
to obtain a lower limit on the ionic column density ($N_{ion}$), if a 
\emph{lower limit} on the EW of the absorption feature is 
available. Such a lower limit on $N_{ion}$ is valid for any value of the 
velocity width (b) of the absorber. Where no lower limit on 
the EW exists, absorption is not significant (at 90$\%$ 
confidence). However, in this case, for an assumed b value, we can use 
the \emph{upper limit} on the EW to get an upper limit 
on $N_{ion}$. In such cases, we assumed a velocity width of 
$b \sim 100$ km $\rm{s}^{-1}$, since this is roughly the smallest 
width of a feature that the MEG can resolve although it 
is considerably larger than the average value of 
$<b>= 40 \pm 13$ km $\rm{s}^{-1}$ found by \citet{b20} in 
signatures of local \oxysix absorption.

\section{Spectral Fitting}
\label{sec:soft}
  We used XSPEC v.11.2.0 for spectral fitting to the MEG spectra. All
spectral fitting was carried out based on the best-fitting continuum 
models from \citep{b15}. Spectral fitting was carried out in the 
0.5-5~keV energy band, excluding the 2.0-2.5 keV region, which suffers 
from systematics as large as $\sim 20\%$ in the effective area due 
to limitations in the calibration of the X-ray telescope
\footnote{http://asc.harvard.edu/udocs/docs/POG/MPOG/node13.html}. 
 We analyzed data binned at $\sim 0.02$\AA, which is approximately the
         MEG FWHM spectral resolution ($0.023$\AA). This MEG spectral 
resolution corresponds to FWHM velocities of $\sim 280$ and $560 \rm km \
   s^{-1}$ at observed energies of 0.5 and 1.0 keV respectively. We used the
C-statistic \citep{b3} for finding best-fit model parameters and quote 
90$\%$ confidence, one-parameter errors.

We proceeded to fit the MEG spectra for the discrete absorption 
transitions in Table~\ref{tab:3} by adding an inverted Gaussian model 
component to the best-fitting continuum models detailed in 
\citet{b15}. The width of the inverted Gaussian was chosen to be $>100$ km 
$\rm{s}^{-1}$, which is approximately the lower limit of the instrumental 
velocity resolution. We fixed the redshift of the Gaussian 
components at $z=0$ and allowed the rest-energy of the component to vary 
by $\pm 1200$ km $\rm{s}^{-1}$ from the rest-frame energies of the 
transitions listed in Table~\ref{tab:3}. The allowed velocity range 
is identical to that used by \citet{b20} and \citet{b14} 
in searches for highly ionized Oxygen absorption in the vicinity 
of the Milky Way. 

\section{Results}
\label{sec:results}
Of the fifteen AGN sightlines in our sample, only the sightlines to NGC 4051 
and MCG-6-30-15 exhibit absorption features within 
$\pm 1200$ km $\rm{s}^{-1}$ of their rest-frame energy at $z=0$ at 
$\geq 99\%$ confidence ($\Delta C \geq 11.3$ for three additional 
parameters). The sightlines to F9, NGC 4593, NGC 3227 and MCG-6-30-15 exhibit 
absorption features within $\pm 1200$ km $\rm{s}^{-1}$ of their 
rest-frame energy at $z=0$ at $\geq 90\%$ but $\leq 99\%$ confidence. 
Of the metals we searched for, only Ne, Mg and Si absorption signatures were 
detected at $>90\%$ confidence. 

\begin{table*}
\centering
\begin{minipage}{135mm}
\caption{Results of spectral fitting for absorption due to hot Ne within 
$\pm 1200$ km $\rm{s}^{-1}$ from transition rest-energies at $z=0$ 
\label{tab:4}. Of the four absorption features listed here, only one 
is likely to be mostly due to hot local gas (see text for discussion). 
Columns 2 and 3 show the best-fit EW for \neniner and \nelya respectively 
and (in brackets) 
the improvement in the fit-statistic upon the addition of the inverted 
Gaussian model component to the continuum. Columns 4 and 5 show the ionic 
column densities of \nenine and \neten respectively as estimated from a 
curve-of-growth analysis (see text for details). Column 6 shows 
the velocity centroid offset from z=0 (LSR) of the \neniner and \nelya 
features respectively. (log) Column densities 
are rounded to the nearest 0.05.}
\begin{tabular}{@{}lrrrrrr@{}}
\hline
Sightline& EW(\neniner) &EW(\nelya) &$\rm{N}_{\rm{Ne~\sc{IX}}}$ &$\rm{N}_{\rm{Ne~\sc{X}}}$ &v(Ne~\sc{IX}) &v(Ne~\sc{X}) \\
 & (eV)($\Delta$C) & (eV)($\Delta$C) & ($\rm{cm}^{-2}$) 
& ($\rm{cm}^{-2}$)
& (km $\rm{s}^{-1}$) & (km $\rm{s}^{-1}$) \\
\hline
NGC 4051 &$2.64^{+0.79}_{-0.95}(17.6)$ &$1.14^{+0.54}_{-0.66}(7.2)$ &$16.60\pm0.15$  & $16.45^{+0.20}_{-0.40}$& $+420^{+520}_{-450}$&$-175^{+205}_{-295}$ \\
MCG-6-30-15 &$0.89^{+0.38}_{-0.50}(8.4)$ &$1.03^{+0.59}_{-0.48}(11.3)$ &$16.05^{+0.20}_{-0.25}$ &$16.40^{+0.25}_{-0.35}$ &$-65^{+260}_{-165}$ &$615^{+295}_{-235}$ \\
\hline
\end{tabular}
\end{minipage}
\end{table*}

Tables~\ref{tab:4} shows the best-fitting model parameters 
for the strongest detections of \neniner and \nelya absorption features. The 
strongest absorption signatures lie along the sightlines to NGC~4051 and 
MCG-6-30-15 respectively, where at least one of the Ne absorption features 
has been detected at $\geq 99\%$ confidence. Table~\ref{tab:5} similarly 
details the best-fit model parameters from a detection of \silya at 
$>3\sigma$ significance and a detection of \mglya at $>90\%$ confidence 
along the sightline to MCG-6-30-15. Listed in 
Tables~\ref{tab:4} and \ref{tab:5} are the equivalent widths 
(EW) of the absorption features and (in brackets) the improvement in the 
C-statistic upon addition of the inverted Gaussian model component to the 
continuum. Also listed in Tables~\ref{tab:4} and \ref{tab:5} are the 
velocity offsets from $z=0$ of the respective Gaussian centroids and 
limits on the respective ionic column densities as estimated from a 
curve-of-growth analysis as outlined above in \S\ref{sec:obs}.

Although the features listed in Tables~\ref{tab:4} and \ref{tab:5} are 
statistically significant, the kinematics of most of these features 
could be consistent with origin in an AGN outflow. The 
absorption features detected towards NGC~4051 for example, are 
barely kinematically consistent with an origin in gas at $z=0$, or indeed 
with an origin in the same gas (within $90\%$ errors, see 
Table~\ref{tab:4}). Nevertheless, 
at $cz=725$ km $\rm{s}^{-1}$, NGC~4051 is very near and with a \chandra 
gratings energy resolution of FWHM $\sim 500$ km $\rm{s}^{-1}$ at the energy 
of the \neniner transition (0.922 keV), it is kinematically difficult to 
distinguish X-ray absorption intrinsic to the NGC 4051 outflow from 
X-ray absorption due to hot local gas. Some or most of this hot gas may 
be associated with the warm absorbing outflow in this AGN. 

Along the MCG-6-30-15 sightline, only the \neniner absorption feature is 
kinematically consistent with $z=0$ absorption. 
The \nelya absorption feature in Table~\ref{tab:4} (at $+615^{+295}_{-235}$ 
km $\rm{s}^{-1}$), coincides kinematically with the \silya and \mglya 
absorption features in Table~\ref{tab:5} (at $\sim +615$ and $\sim +530$ km 
$\rm{s}^{-1}$), but these features are not kinematically consistent with an 
origin in hot local gas at $z=0$. Thus, we conclude that the \nelya, \silya 
and \mglya features along the sightline to MCG-6-30-15 originate in a 
warm absorber outflowing from the AGN at $-1710$ km $\rm{s}^{-1}$ rather 
than hot local gas (see also \citet{b15}). Therefore, in order to study the 
hot \emph{local} \nelya along the sightline to MCG-6-30-15, we searched 
for upper 
limits on absorption due to \nelya by adding a narrow inverted Gaussian to 
the spectrum at $-65$ km $\rm{s}^{-1}$ from $z=0$ rest-frame energy 
(to compare with the corresponding \neniner absorption). We obtained 
upper limits on absorption due to \mglya and \silya similarly by adding 
a narrow 
inverted Gaussian at the $z=0$ rest-frame energy of the transition. Likewise,
 for NGC~4051, we obtained upper limits on absorption due to \nenine and 
\neten by adding narrow inverted Gaussians at the rest-frame energy of the 
respective absorption transitions at $z=0$. 

\begin{figure*}
\includegraphics[height=5in,width=5in]{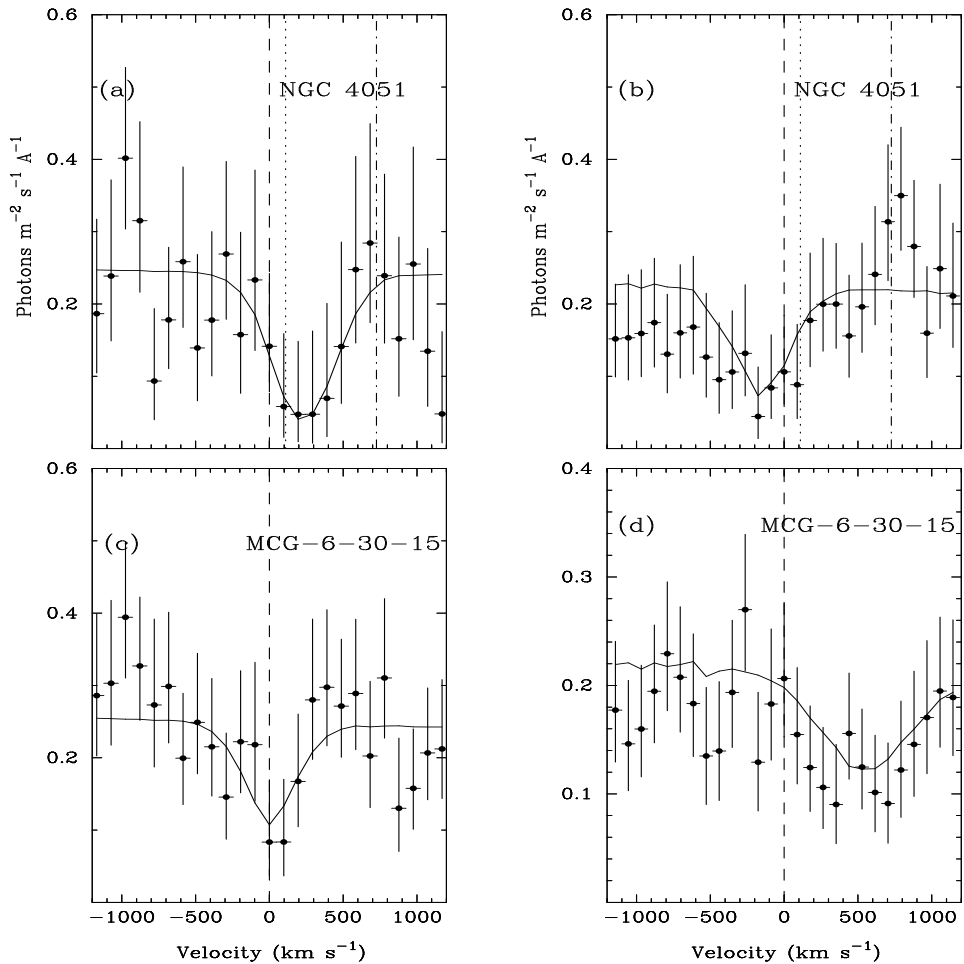}
\caption{Velocity profiles from combined $\pm$1 order \chandra MEG data from 
the AGN in Table~\ref{tab:4}, centered on the LSR 
\neniner transition energy (0.9220 keV) in the left column panels and 
on the \nelya transition energy (1.0218 keV) in the right column panels 
(dashed lines). A positive velocity indicates a redshift relative to these 
energies. The velocity spectra data have been uniformly binned at 0.3eV 
which is approximately the limit of the MEG resolution (0.3eV is the FWHM 
resolution at $\sim 0.4$ keV or about 1/5th the FWHM at the \neniner 
transition energy) . Vertical dotted 
lines in panels (a) and (b) indicate the weighted mean offset velocity 
$cz=+110\pm 115$ km $\rm{s}^{-1}$ of absorption due to \oxyseven and 
\oxyeight (from \citet{b14}). Vertical dash-dot lines indicate the 
rest-frame of NGC~4051 (cz=726 km $\rm{s}^{-1}$). The rest frame of 
MCG-6-30-15 (cz=2325 km 
$\rm{s}^{-1}$) lies outside the $\pm 1200$ km $\rm{s}^{-1}$ range of 
these panels. Superposed is the best-fit inverted Gaussian absorption line 
model (from Table~\ref{tab:4}) and continuum (horizontal solid line). Of 
the four profiles shown here, only the \neniner feature towards MCG-6-30-15 
is likely to be entirely due to hot, local gas (see text). The features 
towards NGC~4051 are most likely due in large part to a warm absorbing 
outflow from this AGN.}
\label{fig:vel_profiles}
\end{figure*}

Fig.~\ref{fig:vel_profiles} is a multipanel plot showing 
velocity profiles of the \neniner and \nelya absorption features along the 
sightlines to NGC~4051 and MCG-6-30-15. The profiles are centered on the 
\neniner transition energy (0.9220 keV) and the \nelya transition energy 
(1.0218 keV) in the LSR respectively (both energies are denoted by 
vertical dashed lines at 0 km $\rm{s}^{-1}$). The vertical dotted line in 
Fig.~\ref{fig:vel_profiles}(a,b) at $+110$ km $\rm{s}^{-1}$ denotes the 
weighted mean offset velocity of the \oxyseven and \oxyeight absorption 
along this sightline detected by \citet{b14}. The 
vertical dash-dot line in Fig.~\ref{fig:vel_profiles}(a,b) at 
$+725$ km $\rm{s}^{-1}$ denotes the recessional velocity (cz) of 
NGC~4051. Superposed on the data is the best-fit inverted Gaussian 
absorption line model (from Table~\ref{tab:4}) and continuum 
(horizontal solid line). 

Of the less significant absorption features detected (at $\geq 90\%$ 
confidence 
but $<99\%$ confidence), those along the sightlines towards F9 and NGC~4593 
are kinematically coincident with their rest-energies at $z=0$ and are 
therefore likely to correspond to hot local gas. A \silya feature detected 
at an offset velocity of $-525^{+450}_{-600}$ km $\rm{s}^{-1}$ along 
the sightline towards NGC~3227 
is more likely to kinematically correspond to an outflow at 
$\sim 1700$ km $\rm{s}^{-1}$ from the AGN (cz$=1160$ km $\rm{s}^{-1}$) than 
absorption due to hot, local gas. We obtained upper limits on absorption due 
to \sifourteen along this sightline by fitting a narrow inverted Gaussian to 
the continuum at the rest-energy of \silya at $z=0$.

\begin{table}
\centering
\begin{minipage}{75mm}
\caption{As for Table~\ref{tab:4}, except for \mglya and \silya along the 
sightline to MCG-6-30-15 \label{tab:5}. These 
features are kinematically consistent with the \nelya feature along this 
sightline in Table~\ref{tab:4} above. Therefore these absorption features 
are most likely 
due to a hot absorbing outflow from MCG-6-30-15 at $\sim -1710$ km 
$\rm{s}^{-1}$ (see text).}
\begin{tabular}{@{}lrrr@{}}
\hline
Transition& EW &N &vel \\
 & (eV)($\Delta$C) & ($\rm{cm}^{-2}$) & (km $\rm{s}^{-1}$) \\
\hline
\mglya &$1.56^{+1.10}_{-0.98}(8.2)$ &$16.55^{+0.20}_{-0.40}$ &$+530^{+450}_{-550}$ \\
\silya &$2.96^{+0.98}_{-1.01}(23.1)$ &$16.60^{+0.15}_{-0.20}$ &$+615^{+255}_{-285}$ \\
\hline
\end{tabular}
\end{minipage}
\end{table}

\subsection{Local, hot gas versus AGN outflow?}
\label{sec:ambiguous}
NGC 4051 and MCG-6-30-15 are two of the closest AGN in our sample, at 
recessional velocities of $cz=726$ km $\rm{s}^{-1}$ 
and $cz=2325$ km $\rm{s}^{-1}$ respectively. Since the spectra of Type I AGN 
typically exhibit hot gas outflowing at several hundred km $\rm{s}^{-1}$, 
the proximity of these two AGN, raises the important issue of confusion 
between absorption due to hot gas at $z=0$ and that due to a warm absorber 
(see also the discussion in \citet{b14}). The proximity of NGC~4051 
and MCG-6-30-15 (and NGC~3227) and the limited gratings spectral resolution 
makes it difficult to distinguish X-ray absorption features along these 
sightlines due to hot local gas from those due to warm absorbing outflows. 
On the other hand, the more distant the AGN, 
the more likely it is that absorption signatures at offset velocities very 
close to their $z=0$ rest-frame energy \emph{are} due to hot, 
local gas. The relation in Figure~3 of 
\citet{b14} shows this effect quite dramatically. Thus, outflows previously 
thought to be associated with more distant AGN, e.g. PG~1211+143 
\citep{b32} and PDS~456 \citep{b33} are kinematically much more likely to 
correspond to hot, local gas. We note that the surprisingly high Fe 
column densities towards both 
of these AGN might be accounted for by the intersection of both of these 
sightlines with the limb of the local Northern Polar Spur (NPS) structure. 
A third sightline through the NPS, towards MCG-6-30-15 also exhibits 
strong absorption due to a large column of highly ionized Fe that is 
kinematically consistent with a local origin \citep{b35}. The NPS is a 
local feature believed to correspond to the superposition of 
several supernovae and is clearly seen in X-ray emission maps (see e.g. 
\citet{b46}) and in maps of polarization towards nearby stars \citep{b49,b47,
b48}. An alternative hypothesis is that the NPS may actually be a 
much larger feature subtended at the Galactic center \citet{b51,b50}, 
although we shall not consider this hypothesis further here. If 
the progenitor supernovae of the NPS were rich in pure Fe, this could 
account for the anomalous column of Fe along these sightlines. We intend to 
investigate this possibility in future work.

\subsection{Column densities of hot local metals}
\label{sec:columns}
Table~\ref{tab:6} lists the limits on the ionic column densities for all 
the ions from spectral fitting. Most of the results in Table~\ref{tab:6} 
are upper limits, indicating that discrete absorption features due to 
local gas are not detected at $>90\%$ confidence in most cases. None 
of the sightlines exhibited absorption due to local N at $>90\%$ confidence. 
The spectra of three AGN, namely NGC~3227, NGC~3516 and 
NGC~7314, were too heavily absorbed at $\leq 0.8$ keV even to obtain 
meaningful limits on \nseven absorption. We found that assuming different 
values of the velocity width (e.g. $b \sim 50$ or $\sim 200$ 
km $\rm{s}^{-1}$), led to small changes in estimates of $N_{ion}$ in 
Table~\ref{tab:6}, $\log (\Delta N_{ion})<0.2$ for $b=50$ km $\rm{s}^{-1}$ 
and $\log (\Delta N_{ion})<0.1$ for $b=200$ km $\rm{s}^{-1}$ respectively 
using the linear part of the curves-of-growth.

\begin{table*}
\begin{minipage}{135mm}
\caption{Results of spectral fitting for hot gas at $z=0$ as discussed 
in text\label{tab:6}. Columns 2-8 show the logarithm of the ionic 
column densities of \nseven,\nenine,\neten,\mgeleven,\mgtwelve,\sithirteen 
and \sifourteen respectively as estimated from a curve-of-growth 
analysis as described in the text. (log) Column densities are rounded to the 
nearest 0.05. Lower limits on ionic column densities 
are valid for all values of $b$. Upper limits 
are valid for an assumed velocity width of $100$ km $\rm{s}^{-1}$ 
(unless specified otherwise) which is approximately the lower bound on 
the instrumental resolution. A choice 
of $b=100$ km $\rm{s}^{-1}$ is larger than the values inferred by 
S03 for local \oxysix absorption features. $^{a}$ The soft X-ray spectrum 
is too absorbed to obtain limits. }
\begin{tabular}{@{}lrrrrrrr@{}}
\hline
Sightline&$\rm{(log)N}_{\rm{N~\sc{VII}}}$ &$\rm{N}_{\rm{Ne~\sc{IX}}}$ &$\rm{N}_{\rm{Ne~\sc{X}}}$   &$\rm{N}_{\rm{Mg~\sc{XI}}}$ &$\rm{N}_{\rm{Mg~\sc{XII}}}$ &$\rm{N}_{\rm{Si~\sc{XIII}}}$ &$\rm{N}_{\rm{Si~\sc{XIV}}}$ \\
 & ($\rm{cm}^{-2}$)& ($\rm{cm}^{-2}$)& ($\rm{cm}^{-2}$)& ($\rm{cm}^{-2}$)& ($\rm{cm}^{-2}$) & ($\rm{cm}^{-2}$)& ($\rm{cm}^{-2}$)\\
\hline
F9 &$<15.30$ &$<17.00$ &$16.35^{+0.25}_{-0.35}$ &$<16.15$  & $<16.60$ &$<17.30$ & $<16.10$\\
3C 120 &$<18.60$ &$<16.90$ & $<15.85$  & $<15.70$ & $<16.75$ &$<16.65$ &$<16.20$ \\
NGC 3227 &$^{a}$ &$<16.40$ &$<16.70$ &$<18.95$ &$<16.90$ &$<17.75$ & $<17.10$\\
NGC 3516 &$^{a}$ &$<16.70$ &$<16.95$ &$<16.15$  &$<16.30$ &$<17.35$ & $<16.70$\\
NGC 3783 &$<15.80$ &$<15.80$ &$<14.95$ & $<15.75$ &$<15.20$ &$<14.75$ &$<14.90$ \\
NGC 4051 &$<16.70$ &$<15.90$ &$<16.80$  & $<16.65$& $<16.60$ & $<17.75$& $<16.70$\\
Mkn 766 &$<16.95$ &$<15.85$ &$<16.10$ &$<16.40$ &$<16.35$ &$<17.30$ & $<16.25$\\
NGC 4593 &$<16.65$ &$16.05^{+0.25}_{-0.80}$ &$<16.10$ & $<16.20$& $<16.15$ & $<16.35$& $<16.45$\\
MCG-6-30-15 &$<16.65$ &$16.05^{+0.20}_{-0.25}$ &$<16.25$ &$<18.15$ &$<16.55$ &$<15.95$ & $<16.50$\\
IC 4329A &$<17.70$ &$<16.50$ &$<15.95$ &$<15.50$ &$<16.35$ &$<15.90$ & $<16.25$\\
Mkn 279 &$<16.10$ &$<16.50$ &$<16.45$ &$<16.80$ &$<16.50$ & $<16.05$& $<16.45$\\
NGC 5548 &$<16.40$ &$<16.10$ &$<16.80$ &$<16.40$ & $<16.35$ &$<16.20$ & $<16.35$ \\
Mkn 509 &$<16.45$ &$<16.20$ &$<15.80$ &$<16.10$ &$<16.10$ &$<16.05$ &$<15.80$ \\
NGC 7314 &$^{a}$ &$<17.05$ &$<16.80$ &$<15.90$ &$<16.50$ &$<17.55$ & $<16.45$\\
Akn 564 &$<16.20$ &$<15.85$ &$<15.85$ &$<16.20$ &$<16.35$ &$<16.75$ & $<16.20$\\
\hline
\end{tabular}
\end{minipage}
\end{table*}

\citet{b20} and \citet{b14} associate local, highly 
ionized O absorption with known local structures such as the Magellanic 
Stream (MS), Complex C and Extreme Positive North (EPn), as well as a 
diffuse Local Group (LG) and (potentially) the warm/hot IGM. Several of the 
sightlines in Table~\ref{tab:6} can be identified with these structures: one 
(F9) with the MS, three (NGC 4593, NGC 4051, NGC 3227) are identified 
with EPn and one (MCG-6-30-15) which we associate with the NPS. In the 
southern Galactic hemisphere, the sightline to Akn~564 passes through the 
Magellanic Stream extension (MSe), but this may also be LG \citep{b20}. 

\citet{b20} conclude that most of the high velocity (v $\sim$ 100-400 km 
$\rm{s}^{-1}$) \oxysix gas in the vicinity of the galaxy is created by 
collisional ionization. Furthermore, gas in the low redshift IGM is far 
more likely to be collisionally ionized than photoionized \citep{b42}. 
Therefore, if we assume that the hot gas in the vicinity of the Galaxy 
is in collisional ionization equilibrium (CIE), and that the 
absorption signatures discussed here and in \citet{b14} 
are due to the same gas, it is possible to establish temperature constraints 
on the gas, whether it is local to our Galaxy or low redshift WHIGM. 
\citet{b21} calculate 
$\rm{N}_{\rm{Ne~IX}}$/$\rm{N}_{\rm{Ne~X}}$ and 
$\rm{N}_{\rm{Si~XIII}}$/$\rm{N}_{\rm{Si~XIV}}$ for gas in CIE, so where there
 are statistically significant lower limits on an ionic column along a 
given sightline, we can constrain the temperature of local gas along 
that sightline with some confidence. However, from searching for seven 
absorption transitions along each of fifteen different sightlines through 
the hot, local gas, there are only three \emph{lower} limits on ionic 
column densities. The corresponding temperature limits (assuming CIE) 
using the $\rm{N}_{\rm{Ne~IX}}$/$\rm{N}_{\rm{Ne~X}}$ ratio are: 
$T<10^{6.75}$ K towards MCG-6-30-15, $T<10^{6.70}$ K towards NGC 4593 and 
$T>10^{6.35}$ K towards F9. The sightline to F9 provides a \emph{lower} 
limit on the temperature because we detect \nelya and not \neniner.

\section{Comparison with signatures of Oxygen absorption}
\label{sec:compare}

\begin{figure*}
\includegraphics[height=4in,width=5in]{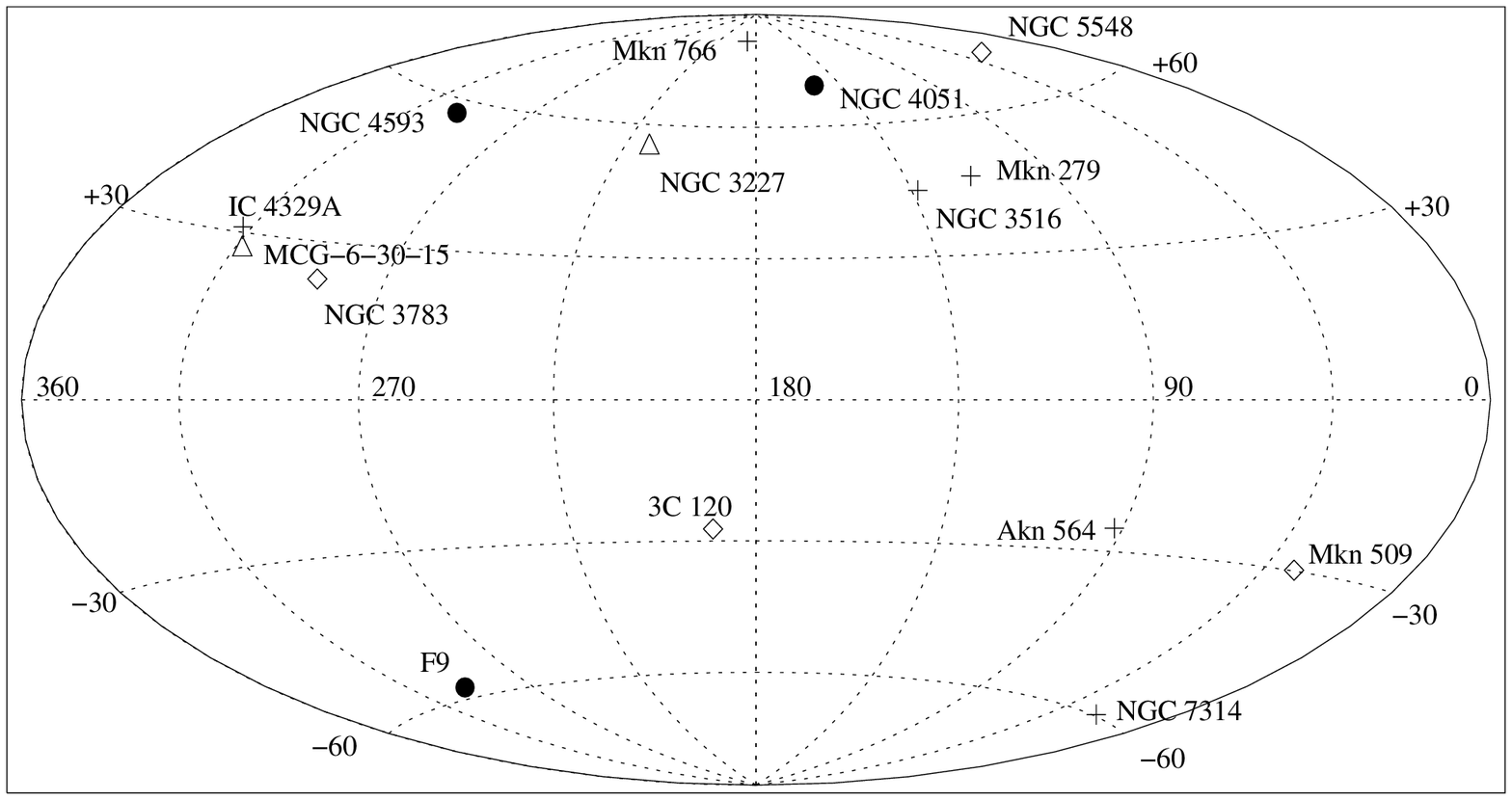}
\caption{All-sky Hammer-Aitoff projection of the sightlines to the 
15 AGN in this study. In this projection, the Galactic anticenter is at the 
center of the figure and Galactic longitude increases to the left. 
Crosses indicate nondetection of local, highly ionized Ne, Mg, Si or O 
along the sightline (measurements of O from \citet{b14}). Diamonds 
indicate non-detection of local, highly ionized Ne or Si, but detection of 
highly ionized O along the sightline. Triangles indicate detection of 
local, highly ionized Ne or Si along the sightlines at $>90\%$ confidence, 
but no corresponding absorption due to local, highly ionized O. 
Filled-in circles denote detection of local, highly ionized Ne or Si a
long the sightlines at $>90\%$ confidence \emph{plus} corresponding 
local, highly ionized O absorption.}
\label{fig:allsky_map}
\end{figure*}

\citet{b14} showed that seven of the fifteen sightlines in our 
sample exhibit discrete 
absorption features due to local \oxyseven and \oxyeight. 
\citet{b20} found local \oxysix ($\lambda 1031.926$ \AA) 
absorption along 59 of 102 sightlines towards UV bright AGN/QSOs at 
high Galactic latitudes ($|b| \geq 30^{\circ}$). Several of the 59 
sightlines coincide with sightlines in our sample (F9, 
NGC~5448, Mkn~509, Akn 564), of which three (F9, Mkn 509, Akn 564) show 
significant \oxysix absorption. In Figure~\ref{fig:allsky_map}, we show 
the sightlines to the fifteen AGN in our sample in Hammer-Aitoff 
projection. Crosses indicate non-detection of \emph{any} 
highly ionized local gas (Ne, Mg, Si and O) along the sightline. Triangles 
indicate detection (at $>90\%$ confidence) of highly ionized local gas 
(Ne or Si) with no corresponding highly ionized O. Diamonds indicate 
the presence of local, highly ionized O along the sightline, but 
non-detection of local, highly ionized Ne, Mg and Si. 
Filled-in circles indicate sightlines along which highly ionized Ne 
or Si has been detected (at $>90\%$ confidence) \emph{and} which 
show absorption due to local, highly ionized O. 

Of the three sightlines which exhibit absorption by local, highly 
ionized Ne at $>90\%$ confidence (listed in Table~\ref{tab:6}), 
two sightlines (F9 and NGC~4593) also exhibit absorption by \oxyseven and/or 
\oxyeight at $>90\%$ confidence \citep{b14}. The sightline 
towards F9 also exhibits \oxysix absorption \citep{b20}. The  
sightlines to MCG-6-30-15 does not exhibit absorption 
due to highly ionized Oxygen. In \citet{b14} we constrained the 
temperature in local gas from limits on $N_{O VIII, VII, VI}$ and an 
assumption of CIE. Towards F9, we found $10^{5.75}<T<10^{6.35}$K. This 
is marginally inconsistent with our estimate of $T>10^{6.35}$K using 
N(\nenine, \neten) in \S\ref{sec:results} above. However, the condition that 
$T<10^{6.35}$K along this sightline is derived assuming 
a b-parameter of precisely $100$ km $\rm{s}^{-1}$ using $N_{O~VIII}$ in 
a curve-of-growth analysis. A choice of a slightly larger b-parameter (which 
would be allowed by the data) would yield a temperature upper limits
 compatible with that derived using N (\nenine, \neten). The combined results
using N(\oxysix, \oxyseven, \oxyeight, \nenine \& \neten) along the sightline
 to F9, suggests that the temperature in the local hot gas in the MS is 
close to $T\sim 10^{6.35}$ K and that there is some velocity broadening 
of the hot gas along this sightline. The temperature of the gas along the 
sightlines to NGC 5448 and Akn 564 was also 
constrained by \citet{b14} to be $T>10^{6.2}$ K and $T<10^{6.1}$K 
respectively. However, the absence of significant absorption along 
the sightlines to NGC 5548 and Akn 564 in this study (see 
Table~\ref{tab:6}), means that we cannot provide additional constraints 
on the temperature of gas along these sightlines.

\section{Possible constraints on metallicities in hot local gas}
\label{sec:metals}
The temperature ranges derived from the ratios of column densities in 
\S\ref{sec:results} above and by \citep{b20} and \citep{b14} 
suggest that we can assume that most of the O, Ne, Mg and Si in the 
hot gas in the vicinity of the Galaxy is in H-like and He-like ions. 
If we assume that the gas is in CIE, then for a range of 
temperatures, such as those discussed here, the fraction in H-like 
or He-like ions is much larger than in other ionic states \citep{b21} 
(e.g. in the case of Oxygen ions, \oxyeight / \oxynine $\gg$ 1, 
\oxyseven / \oxysix $\gg$ 1). Thus, in a simple approximation, the 
largest upper (or lower) limit on the column density of the 
He-like or H-like ions of a particular element is approximately the 
upper (lower) bound on the amount of the element present. 

Therefore, given a lower limit to the column density N(X) of an 
element X, using this simple approximation, we can use upper limits on 
the column density N(Y) of element Y to obtain a lower limit 
on N(X)/N(Y) along a sightline. Limits on relative metal abundances 
in the hot local gas, allow us to compare the results with solar relative 
abundances (see Table~\ref{tab:2}) 
and thereby establish limits on the composition of the hot gas. The
 relative depletion or enrichment of certain metals can provide 
clues to the environment, formation and evolution of the hot gas in local 
structures and the hot IGM. Note that it is not possible to obtain reliable 
upper limits on the ratio N(X)/N(Y), since an upper limit on N(X) is 
potentially much more inaccurate than a lower limit (a firm lower 
bound on the column), due to the possibility of additional columns of 
more or less ionized gas than those considered here.

\begin{table}
\begin{center}
\begin{minipage}{78mm}
\caption{The (log) abundances of elements relative to Oxygen in the 
hot gas in the vicinity of the Galaxy, assuming CIE. Values 
in (brackets) correspond to a solar Ne abundance from \citep{b31}.
\label{tab:7}}
\begin{tabular}{@{}lrrrr@{}}
\hline
Sightline & [O/Ne] & [O/N] & [O/Mg] & [O/Si]\\
\hline
$\odot$ & 0.84(0.64) & 0.90 & 1.29 & 1.38\\  
F9 &$>-0.90$ &$>0.80$ & $>-0.50$ & $>-1.20$\\  
3C 120 & $>-1.22$ & $>-2.86$ & $>-1.03$ & $>-0.47$\\  
NGC 3783 &$>0.04$ &$>0.06$ & $>0.08$& $>0.95$\\  
NGC 4051 & $>-0.35$& $>-0.24$ & $>-0.22$ & $>-1.30$\\  
NGC 5548 & $>-0.87$ & $>-0.46$ & $>-0.49$ & $>-0.42$\\  
Mkn 509 & $>-0.93$ & $>-1.16$ & $>-0.85$ &$>-0.77$\\  
\hline
\end{tabular}
\end{minipage}
\end{center}
\end{table}

\begin{table}
\begin{center}
\begin{minipage}{85mm}
\caption{The (log) abundance of Ne relative to O, N, Mg and Si in the 
hot gas in the vicinity of the Galaxy. We assumed CIE in calculating these 
ratios. Values in (brackets) correspond to a solar Ne abundance from 
\citep{b31}. $^{a}$ There is no meaningful upper limit on N(O) for the the 
sightline to F9, since $\log$N(\oxyseven)$>16.1$ (possible saturation) 
and $\log$N(\oxyeight)$<16.6$ \citep{b14}.\label{tab:8}}
\begin{tabular}{@{}lrrrr@{}}
\hline
Sightline & [Ne/O] & [Ne/N] & [Ne/Mg] & [Ne/Si]\\
\hline
$\odot$ & -0.84(-0.64) & 0.04(0.24) & 0.50(0.70) & 0.54(0.74)\\  
F9 & $^{a}$ & $>0.70$ & $>-0.60$ & $>-1.30$\\  
NGC 4593 & $>-1.99$ & $>-1.41$ & $>-0.96$ & $>-1.19$\\  
MCG-6-30-15 & $>-0.85$ & $>-0.60$ & $>-0.15$ &$>-0.60$\\  
\hline
\end{tabular}
\end{minipage}
\end{center}
\end{table}

Tables~\ref{tab:7} and \ref{tab:8} list constraints on the 
metal abundances relative to O and Ne respectively, in the hot gas in 
the vicinity of our Galaxy. Clearly the lower limits on the 
relative abundances in the local, hot gas are in general too small to be 
interesting, with a few exceptions. In Table~\ref{tab:7}, the lower 
limit on [O/N] along the sightline to F9 
is close to the solar value, while the other ratios are one or two 
orders of magnitude lower, which may suggest a relative underabundance 
of N in the MS. Also in Table~\ref{tab:7}, the lower limit on [O/Si] 
along the sightline to NGC 3783 is close to the solar ratio, while the 
other ratios are significantly lower, which may be hinting at a relative 
underabundance of Si along this sightline. In Table~\ref{tab:8}, [Ne/N]
 along the sightline to F9 \emph{exceeds} the solar value by 
$\sim 3.9$, and [Ne/Si] along the sightline to \mcg is roughly solar. 
The other Ne ratios are one or two orders of magnitude below their solar 
values. This again suggests an underabundance of N in the hot gas in the MS. 

\section{Conclusions}
\label{sec:conclusions}
Half of the baryonic matter in the local universe seems to be missing. The 
search for the hottest component of the missing matter has only been possible 
with the latest generation of high spectral resolution X-ray telescopes. 
Hot gas in the vicinity of the Galaxy may be due to local WHIGM or it 
may reside either in a hot Galactic halo or locally in a thick disk and 
has only recently begun to be studied in the 
X-ray band. We assembled a small sample of type~I AGN observed with the
high resolution X-ray gratings on board \chandra and we have applied 
a uniform analysis to detect soft X-ray absorption by hot gas in the 
vicinity of our Galaxy. This study is an extension of 
our previous study of \oxyseven and \oxyeight absorption by hot local gas 
\citep{b14}. 

Three of the fifteen sightlines in our sample (towards F9, 
NGC 4593 and MCG-6-30-15 respectively) exhibit \neniner or \nelya 
absorption due to hot, local gas at $\geq 90 \%$ confidence. We identify 
these absorption features with the hot phase of local structures. Such 
local structures are either in the disk of the Galaxy (e.g. the Local 
Bubble or superbubbles such as the NPS) or lie 
above the disk in HVCs or other structures, such as a Galactic halo. Hot 
gas in a thick disk is expected to have low offset velocity from $z=0$ 
(typically $\leq 100$ km $\rm{s}^{-1}$) and hot gas above the disk in 
HVCs or in a Galactic halo is expected to have a higher offset velocity 
from $z=0$ (typically $100-400$ km $\rm{s}^{-1}$). The 
sightlines to the AGN in our sample lie well away from the Galactic 
plane ($b \geq 30^{\circ}$) and \chandra does not possess the velocity 
resolution to distinguish between kinematic signatures of a thick disk or 
Galactic halo. Therefore we cannot tell whether absorption is local to 
the disk, or further out in a halo, where there are no known local 
structures along a sightline. Absorption studies in the UV band with 
\fuse indicate that there is a thick Galactic disk of \oxysix, with a scale 
height of $\sim 2.3$ kpc, as well as a patchy overdensity of \oxysix in 
the northern Galactic hemisphere from $b \sim 45^{\circ}$ to $90^{\circ}$ due 
to the Local Bubble and superbubbles (including the NPS) \citep{b43}. X-ray 
spectral studies of Galactic X-ray binaries confirm that there is a hot, 
thick disk with scale height of $\sim 1-2$ kpc associated with our Galaxy 
(see e.g. \citet{b44,b45}). The column densities inferred from 
\citet{b44,b45} are consistent with the limits inferred here, so it may be 
that most or all of the hot gas locally is associated with a Galactic 'thick 
disk' of hot gas, with enrichment along particular sightlines in the 
northern Galactic hemisphere due to local superbubbles such as the NPS.

There can be considerable ambiguity in distinguishing local hot absorbing 
gas and hot absorbing outflows from AGN, especially with the limited spectral
 resolution of the \chandra gratings. This is particularly true of some 
of the AGN in the present study. Two sightlines in our study (towards 
MCG-6-30-15 and NGC 4051 respectively) exhibit 
absorption features within $\pm 1200$ km $\rm{s}^{-1}$ of their 
rest-energies at $z=0$ at confidence levels $>99\%$. However, the kinematics 
of the absorption signatures suggest that these features are most 
likely due to absorption in hot gas outflowing from the respective 
background AGN. Nevertheless, the more distant the AGN, the more likely that 
absorption signatures in the AGN spectrum which kinematically coincide with 
the AGN recession velocity are due to hot, local gas. This is demonstrated 
by the relation in Fig.~3 of \citep{b14}. Two of the AGN in Fig.~3 of 
\citep{b14}, namely PG~1211+143 and PDS~456, also exhibit absorption due to 
a very large local column of Fe, which is surprising. We point out that 
absorption by the local NPS bubble, if Fe-enriched, could account for the 
very large column density of Fe along the sightline to these AGN, and 
towards MCG-6-30-15 as well. Thus, hot gas in the disk of the Galaxy may 
account for much of the $z=0$ absorption in the soft X-ray spectra 
of several AGN.

We assumed collisionally ionized equilibrium (CIE) to use limits on ionic 
column densities to derive limits on the temperatures in the hot, 
local gas. We find that in the southern Galactic hemisphere, the sightline 
to F9 through the MS yields $T>10^{6.35}$K, which is marginally consistent 
with our temperature limits from N(\oxyseven)/N(\oxyeight) \citep{b14}. 
Likewise, in the northern Galactic hemisphere we find temperature limits 
of $T<10^{6.75}$ K towards MCG-6-30-15 and $T<10^{6.70}$ K towards NGC 4593. 
For a range of temperatures, such as those discussed 
here, derived limits on the H- and He-like ion column densities allow us to 
establish constraints on the 
relative abundances of N, Ne, Mg and Si (as well as O from \citet{b14}) 
along the sightlines in our sample. In general, we found lower 
limits on the relative abundances of the metals that are one or two orders 
of magnitude below the corresponding solar values (see Tables~\ref{tab:7}
and \ref{tab:8}). However, we do find evidence for a 
relative \emph{underabundance} of N in the hot gas in the MS along the 
sightline towards F9. 

The question of whether the hot gas detected in the vicinity of the 
Galaxy is in part (or at all) associated with the missing matter remains as 
yet unanswered. The environment of the Galaxy is complex, consisting of 
satellite galaxies (Magellanic Clouds), tidal trails (the Magellanic 
Stream), infalling HVCs and outflows from supernovae in the Galaxy. 
Standard theories of structure formation in the universe suggest that our 
Local Group (our Galaxy and M31) should have formed from a low density 
filament of gas, the heated remnants of which should persist today. 
The local filament should be metal-enriched by outflows from structures that 
formed within it. However, the upper limits on column densities that we 
derive here and in \citet{b14} seem to be consistent with estimates of 
column densities from studies of Galactic halos in edge-on normal spiral 
galaxies \citep{b25}. Thus, our first systematic (but hardly comprehensive) 
look at the hot, X-ray absorbing component of gas in the vicinity of our 
Galaxy suggests that our Galaxy is embedded in a hot halo or hot 
thick disk. 

\section*{Acknowledgments}
We gratefully acknowledge support from NSF grant AST0205990 (BM) and NASA 
grant AR4-5009X issued by CXC operated by SAO under 
NASA contract NAS8-39073 (TY). We made use of the HEASARC on-line data 
archive services, supported by NASA/GSFC and also of the NASA/IPAC 
Extragalactic Database (NED), operated by the Jet Propulsion Laboratory, 
CalTech, under contract with NASA. Thanks to the \chandra instrument 
and operations teams for making the observations possible. Thanks to 
M. Coleman Miller, David Strickland and Andy Fabian for useful 
discussions and thanks to Ken Sembach for bringing to our attention 
the uncertainty in the solar Ne abundance. Thanks to the anonymous referee 
for their comments which helped improve this paper.

\bsp

\label{lastpage}

\end{document}